# Temperature-independent optical cavities for laser frequency stabilization

**NIKITA O. ZHADNOV,[1] ANATOLY V. MASALOV[1,2]**

[1] *P. N. Lebedev Physical Institute, Russian Academy of Sciences, Leninskiy prospect 53, 119991 Moscow, Russia*
[2] *National Research Nuclear University MEPhI, Kashirskoe shosse 31, 115409, Moscow, Russia*
*nik.zhadnov@yandex.ru*

**Abstract:** We propose a method for thermal expansion compensation of reference monolithic optical cavities for laser frequency stabilization. Two schemes of optical cavities are considered: a Fabry-Perot interferometer with a crimp ring and a whispering-gallery-mode cavity with a clamp. In each scheme, thermal expansion compensation is achieved due to the strained connection of the cavity with an element made of a material with a high coefficient of thermal expansion. The temperature region of the cavities' optical length stabilization is estimated.

## 1. Introduction

Materials with low or zero thermal expansion are in demand for the manufacture of high-precision instruments designed to measure physical quantities or create their standards. The limiting characteristics of such equipment are defined by the temperature dependence of the properties of structural materials. In this regard, the search for and creation of materials with zero or negative coefficient of thermal expansion (CTE) is an essential task. These properties are rare in nature. Such behavior was found in some solids: complex metal oxides, zeolites, polymers [1].

The problem of thermal deformations is especially critical when creating reference monolithic optical cavities used to stabilize laser radiation frequency in optical clocks [2,3]. The frequency stability of a Fabry-Perot cavity mode is given by the stability of its length, which in turn depends on temperature. The change in the resonant frequency of the interferometer with temperature has the scale:

$$\delta\nu = -\nu\alpha\delta T, \quad (1)$$

where $\nu = m/(2L/c)$ is the resonant frequency in Hz, $m$ is the resonance order, $\alpha$ is the linear thermal expansion coefficient of the cavity material. So, for a Fabry-Perot cavity with a fused silica body ($\alpha = 5.5 \cdot 10^{-7}$ 1/°C), the resonant frequency deviation will be hundreds of MHz per degree. With an achievable temperature stabilization of $10^{-4}$ °C [4], the frequency instability turns out to be many orders of magnitude greater than the requirements of modern devices. The temperature-compensated glass ULE (Corning Inc.) is the most widely used material for the manufacture of reference ultrastable cavities [3–6]. The convenience of using this glass is primarily because the temperature at which its coefficient of thermal expansion becomes zero lies near room temperature. In addition, it is transparent, has a high Young's modulus, and is not difficult to machine. The dependence of the CTE of ULE glass on temperature near $T_0$ is described by the relation $\alpha_{ULE}(T) = a(T - T_0) + b(T - T_0)^2$ [7], where the linear temperature coefficient is $a \approx 1.6 \cdot 10^{-9}$ 1/°C$^2$ and the quadratic coefficient $b \approx -10^{-11}$ 1/°C$^3$ (the actual values of the coefficients and the value of $T_0$ depend on the brand of glass and are determined by the conditions during its fabrication). It is easy to conclude that the deviation of the cavity temperature by $1\ mK$ from $T_0$ will lead to a relative length change equal to $10^{-15}$, which is acceptable. A negative property of ULE glass is the recrystallization process, which changes the length of the cavity over time and creates a frequency drift of eigenmodes at the level of fractions of a hertz per second [8,9]. To date, in the manufacture of reference cavities, crystalline materials free from recrystallization have gained an advantage, for example, single-crystal silicon [10], which also has "zero points" of CTE at $17\ K$ and $124\ K$. The use of silicon cavities to stabilize laser frequencies with low operating temperatures is acceptable for space-based instruments, while for ground-based equipment this creates additional technical inconveniences.

Several applications of laser systems with reference optical cavities do not require as high frequency stability as optical clocks. Examples include lasers for cooling atoms, precision spectroscopy, optical communication lines, etc. In these cases, in addition to Fabry-Perot cavities, whispering-gallery-mode cavities are used [11–13]. The temperature stability of the mode frequency is one of the main limiting factors of frequency stability for such cavities.

Attempts to compensate thermal expansion of Fabry-Perot etalons [14] and microcavities [15–17] have been made several times. The authors of [16,17] bonded Zerodur layers (a glass-ceramic material with a negative CTE) with a $MgF_2$ cavity with a whispering gallery mode; effective CTE was reduced by more than 6 times.

In this paper, we propose schemes for strained connection of optical cavities with a metal frame, where the thermal expansion of the cavity material is compensated when the CTE of the metal is sufficiently higher than the CTE of the resonator [18]. Section 2 describes an assembly that compensates axial thermal expansion of a Fabry-Perot interferometer, and Section 3 describes a circuit that compensates thermal expansion of a whispering gallery mode cavity. The development of a temperature-independent optical cavity is a step towards compactization, simplification, and cost reduction of laser frequency stabilization systems.

## 2. Monolithic Fabry-Perot cavity

A cavity of this type consists of a cylindrical body with a through-hole along the axis, which is plugged on both sides by mirrors (Fig. 1). With a temperature change, the body material changes the length L of the cavity, and the selected resonant frequency ν changes its value. The resonant frequency drift with a change in temperature T is estimated by the formula (1).

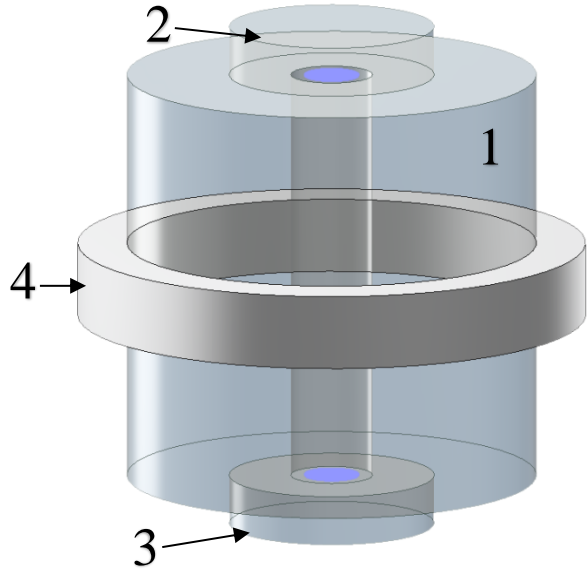

*Fig. 1. Schematic view of a reference Fabry-Pero cavity with a compensation ring. 1 - body, 2,3 - mirrors, 4 - compensation ring.*

For temperature stabilization of the cavity length, we propose to use compression of the body along the diameter with a metal (or other material) ring, the thermal expansion coefficient $\alpha_M$ of which is greater than that of the body material $\alpha$: $\alpha_M > \alpha$ (Fig. 1). When the body of the cavity is compressed with a metal ring, the temperature dependence of the distance between the mirrors changes. If the coefficient of thermal expansion of the ring $\alpha_M$ is greater than $\alpha$, then with the thermal expansion of the body and the ring, the pressure of the ring on the walls of the body decreases, and there is a tendency to reduce the length of the cavity (Fig. 2).

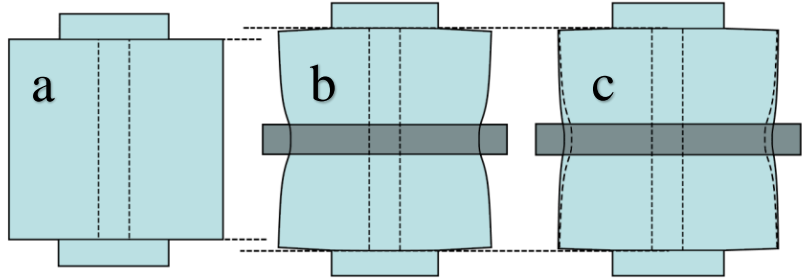

*Fig. 2. Undeformed cavity (a); cavity compressed by the compensating ring (b); cavity with the ring when the temperature rises (c).*

We calculate the thermal expansion of the body with a compensation ring, assuming that the ring is placed on the body in a stressed state. We denote the radii of the body and the ring in the free state (before their connection) $R_0$ and $R_{M0}$: $R_{M0} < R_0$. After the ring is placed on the body, pressure $P_0$ appears on the contact surface, which reduces the body radius (in the contact area) and increases the ring radius. The new body radius will take the value:

$$R \approx R_0 \left(1 - \frac{P_0}{E_M}\right), \qquad (2)$$

and the new radius of the ring is:

$$R_M \approx R_{M0} \left(1 + \frac{P_0}{E_M} \cdot \frac{R_0}{b}\right), \qquad (3)$$

where $E$ and $E_M$ are the Young's modules of the body and the ring, $b$ is the ring's thickness. The formula (2) estimates the effect of compression of a volumetric body, and (3) describes the expansion of a thin-walled pipe under the action of pressure from the inside ([19], §7, problem 4). From the equality of these quantities the expression for the pressure force reads:

$$P_0 \approx \frac{R_0 - R_{M0}}{R_0} \left(\frac{R_0/b}{E_M} + \frac{1}{E}\right)^{-1}. \qquad (4)$$

In this case, the length of the cavity will increase by:

$$L(T_0, P_0) \approx L_0 \left(1 + \frac{2\sigma a}{L_0} \cdot \frac{P_0}{E}\right), \qquad (5)$$

where $\sigma$ is Poisson's ratio, $a$ is the width of the ring. In this formula, it is assumed that the length of the body increases only due to the compression area of size $a$.

With a temperature change, the pressure $P$ will change, and the length estimation will take the form:

$$R \approx R_0 \left(1 + \alpha(T - T_0) - \frac{P}{E}\right), \qquad (6)$$

$$R_M \approx R_{M0} \left(1 + \alpha_M(T - T_0) + \frac{P}{E_M} \cdot \frac{R_0}{b}\right). \qquad (7)$$

As before, due to the equality of the radii, we obtain a new pressure value:

$$P \approx P_0 + (\alpha - \alpha_M)(T - T_0) \left(\frac{R_0/b}{E_M} + \frac{1}{E}\right)^{-1}, \qquad (8)$$

and a new cavity length:

$$L(T, P) \approx L_0 \left(1 + \alpha(T - T_0) + \frac{2\sigma a}{L_0} \cdot \frac{P}{E}\right). \qquad (9)$$

The temperature independence condition is reduced to the equality of lengths (5) and (9):

$$\frac{2\sigma a}{L_0} \cdot \frac{P_0}{E} = \alpha(T - T_0) + \frac{2\sigma a}{L_0} \cdot \frac{P}{E}, \tag{10}$$

which, considering (8), gives:

$$\alpha - (\alpha_M - \alpha)\frac{2\sigma a}{L_0 E}\left(\frac{R_0/b}{E_M} + \frac{1}{E}\right)^{-1} = 0. \tag{11}$$

The thermal expansion compensation condition (11) does not depend on the compression force $P_0$. Only the boundary temperature at which compensation is achieved depends on it. Equation (11) is a condition for the size of the ring. As an example, let us assume that the cavity body is made of fused quartz ($\alpha = 0.55 \cdot 10^{-6}$ 1/°C, $E = 72$ GPa, $\sigma = 0.17$) and the ring is made of aluminum ($\alpha_M = 23 \cdot 10^{-6}$ 1/°C, $E_M = 68$ GPa). Then at $R_0 = 25$ mm, $b = 6$ mm from (11) we obtain an estimate of the required dimensions of the ring: $a/L_0 \approx 0.42$, which is quite practical.

When the temperature independence condition (11) is not exactly satisfied, the presented formulas give an estimate of the effective thermal expansion coefficient of the cavity body with a ring:

$$\alpha_{\text{eff}} \approx \alpha - \frac{2\sigma a}{L_0 E}(\alpha_M - \alpha)\left(\frac{R_0/b}{E_M} + \frac{1}{E}\right)^{-1}. \tag{12}$$

Since the obtained analytical formulas are estimations and do not consider the nonlinear nature of thermal expansion, a calculation of the effective CTE of a cavity with a ring was performed using the ANSYS program. The calculation was made for the cavity model shown in Fig. 1 with the following dimensions: body length $L_0 = 50$ mm, body radius $R_0 = 25$ mm, axial hole diameter 10 mm, ring thickness $b = 6$ mm. The optimal width of the ring is determined in the calculation. Fused silica was chosen as the material for the body and mirrors, and aluminum for the compensation ring. Since quartz glass has a low CTE, a lot of metallic structural materials are suitable for the ring material.

To simulate the compression of a quartz body by the ring, its inner diameter was matched with the outer diameter of the body at high temperature (100 °C). Due to the high CTE of aluminum with respect to quartz, stressed compression occurred with decreasing temperature. The computer simulation was aimed to calculate the distance between the mirrors of such a cavity with temperature. To describe the optical contact between the body and the mirrors, the "Bonded" type of contact was used in ANSYS program, which excludes any relative displacements of the points on the surfaces of bonded objects. The connection between the body and the ring was of the "No separation" type, allowing a small mutual displacement in the direction of the surface between the contacted objects only. To calculate the CTE of the cavity at each chosen temperature T, two simulations were carried out, in which the resonator with the ring was cooled from 100 °C to temperatures $T \pm 0.25$°C. In accordance with formula (12), the cavity CTE should depend on the dimensions of the compensation ring. Ring width was chosen $a \approx 8.6$ mm to achieve zero thermal expansion at a temperature of 21.5 °C. The dependence $\alpha_{\text{eff}}(a)$ at different ring widths is shown in Fig. 3. The resulting dependence is linear with a slope of $-7.8 \cdot 10^{-8}$ [(1/°C)/mm]. According to formula (12) the slope is estimated as: $2\sigma \cdot (\alpha - \alpha_M)/L_0(2 + R_0 E/E_M b) \approx -2.8 \cdot 10^{-8}$ [(1/°C)/mm]. The discrepancy between the simulation results and the analytical formula (12) is due to approximate character of the expressions (2-9) and the lack of consideration for the influence of the axial hole and mirrors.

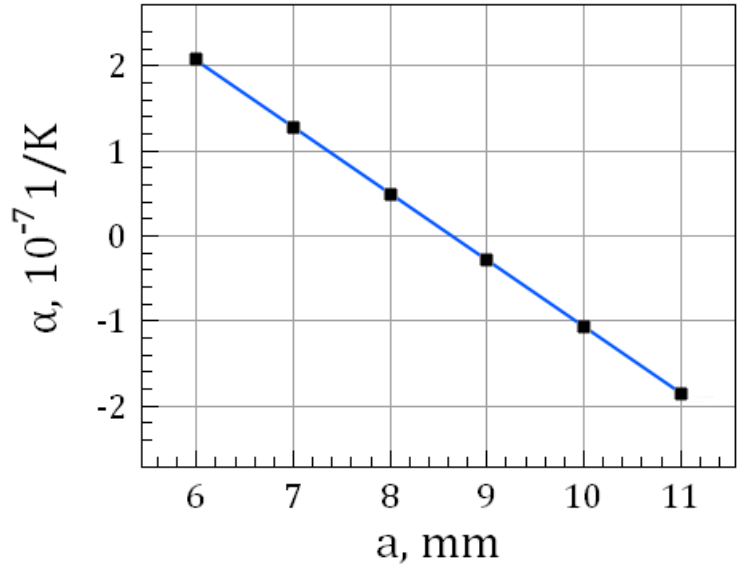

*Fig. 3. A plot of the coefficient of thermal expansion of a fused silica cavity with a ring as a function of the width $a$. CTE was determined in finite element calculation for a cavity temperature 21.5 °C.*

CTE of the cavity with a compensation ring at different temperatures is determined by the temperature dependences of the elastic constants and CTE's of its constituent materials. Such dependences for fused quartz were taken from [20,21] and for aluminum from [22,23]. The modeling of thermal deformations of the studied cavity was carried out for the optimal value of the ring parameters. The resulting dependence of the value $\Delta L/L(T)$ for such a design is shown in Fig. 4. The cavity has a temperature of zero thermal expansion of about 21.5 °$C$ (for this temperature the ring width $a$ was optimized). The resulting linear coefficient of the CTE temperature dependence of the studied cavity model near the zero point is $8.1 \cdot 10^{-10}$ 1/K². For comparison, the specified parameter for ULE thermally compensated glass (Corning part number 7972) is $15.7 \cdot 10^{-10}$ 1/K². The dependence $\Delta L/L(T)$ for a ULE cavity was obtained experimentally in [7]. It should be noted that for different pieces of such glass, even with the same brand, the position of the zero point and temperature coefficients may differ.

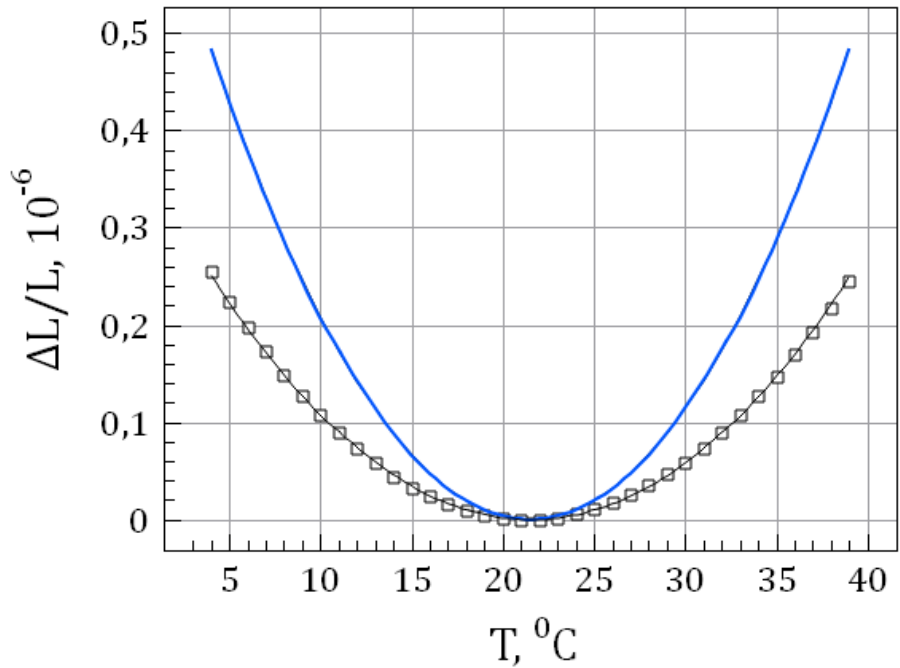

*Fig. 4. Thermal expansion dependences for cavities made of ULE glass (blue curve) and quartz with a compensation ring: black squares are the result of the calculation; black curve is parabolic approximation.*

Surprising that ULE cavities with quartz mirror substrates are characterized by a shift of the zero point to lower temperatures, and a steeper slope of the CTE temperature dependence than for the case with ULE substrates [7].

## 3. Whispering gallery mode cavity

In this section, we consider another type of cavities used to stabilize laser radiation frequency: whispering gallery mode (WGM) resonators [11]. This cavity is a cylinder made of a dielectric material with low optical losses with a diametrical rim, where the WGM can propagate (Fig. 5).

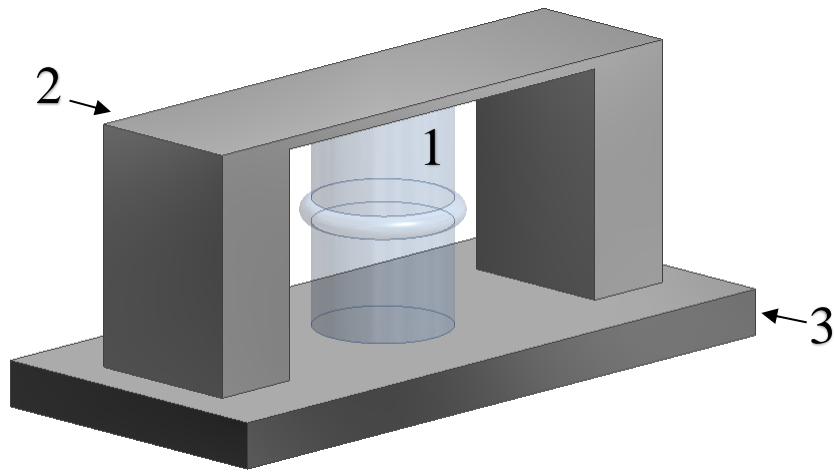

*Fig. 5. The WGM cavity with thermal expansion compensation design. 1 – WGM cavity, 2 – clamp, 3 – base.*

When the temperature changes, the WGM resonant frequency has a shift according to the relation:

$$\frac{d\nu}{dT} = -\nu\left(\frac{1}{nR}\frac{d(nR)}{dT}\right) = -\nu\left(\alpha + \frac{1}{n}\frac{dn}{dT}\right) = -\alpha'\nu, \quad (13)$$

where $\nu = m/(2\pi nR/c)$ is the resonant frequency, $R$ is the radius of the WGM zone, $n$ is the refractive index of the cavity material, and $\alpha' = \alpha + \frac{1}{n}\frac{dn}{dT}$ is the sum of the CTE of the cavity material corrected for the dependence of the refraction on temperature (usually small in comparison with the CTE).

For temperature stabilization of the cavity radius, we propose compressing it with a clamp made of metal or another material, whose thermal expansion coefficient $\alpha_M$ is higher than that of the cylinder material $\alpha$: $\alpha_M > \alpha$. When the clamp is pressed on the cylinder, the temperature dependence of the WGM radius changes. If the coefficient of thermal expansion of the clamp $\alpha_M$ is greater than $\alpha$, then with the thermal expansion of the cylinder and the clamp, the clamp pressure on the cylinder decreases, and there is a tendency to reduce the WGM radius (Fig. 6).

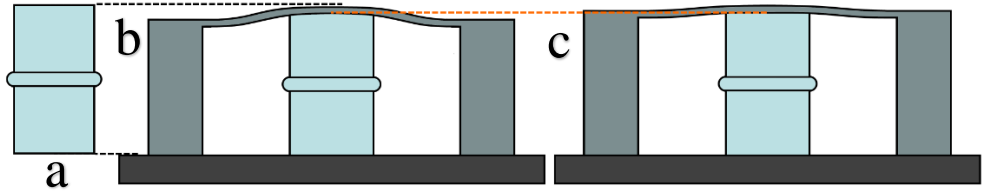

*Fig. 6. Deformations of the WGM resonator; (a) is the view of the cavity before the clamp is installed, (b) is the cavity after the clamp is installed, (c) is the cavity with the clamp when the temperature rises.*

The thermal expansion of the WGM radius of the cylinder with the clamp can be calculated, assuming that the clamp holds the cylinder in a stressed state. We denote the length and the radius of the cylinder in the free state (before fixing the clamp) $L_0$ and $R_0$; length of clamp pillars $L_{M0}$, $L_0 > L_{M0}$. After fixing the clamp on the cylinder, pressure $P_0$ arises, which reduces the length and increases the radius of the cylinder. The new length and radius of the cavity can be estimated as:

$$L \approx L_0\left(1 - \frac{P_0}{E}\right), \quad (14)$$

$$R \approx R_0\left(1 + \frac{\sigma P_0}{E}\right), \quad (15)$$

and for the new length of pillars, together with the crossbar deflection, we have the estimate:

$$L_M \approx L_{M0}\left(1 + \frac{P_0}{E_M}\right) + \frac{P_0 S(d/2)^3}{16 E_M a b^3}, \quad (16)$$

where $S$ is the cross-section of the clamp pillars, $d$ is the length of the clamp crossbar, $a$ and $b$ are the width and thickness of the crossbar. The expression for the crossbar deflection in (16) is a well-known solution to the problem of the deflection of a crossbar with fixed ends and a concentrated force at the center. From the equality of lengths (14) and (16), the crossbar pressure on the cylinder we have:

$$P_0 \approx \frac{L_0 - L_{M0}}{L_0}\left\{\frac{1}{E} + \frac{1}{E_M} + \frac{Sd^3}{128 E_M L_0 a b^3}\right\}^{-1} = \frac{L_0 - L_{M0}}{L_0}\left\{\frac{1}{E} + \frac{1}{E'_M}\right\}^{-1}, \quad (17)$$

where:

$$\frac{1}{E'_M} = \frac{1}{E_M}\left(1 + \frac{Sd^3}{128 L_0 a b^3}\right). \quad (18)$$

The pressure $P_0$ is greater, with greater initial difference in dimensions (14) and (16).

With a change in temperature, the pressure of the clamp $P$ on the cylinder will change and the estimates of lengths will take the form:

$$L \approx L_0 \left(1 + \alpha'(T - T_0) - \frac{P}{E}\right), \tag{19}$$

$$L_M \approx L_{M0}\left(1 + \alpha_M(T - T_0) + \frac{P}{E'_M}\right). \tag{20}$$

Due to their equality, we obtain a new value of pressure:

$$P \approx P_0 + (\alpha' - \alpha_M)(T - T_0)\left(\frac{1}{E} + \frac{1}{E'_M}\right)^{-1} \tag{21}$$

and WGM radius:

$$R \approx R_0\left(1 + \alpha'(T - T_0) + \frac{\sigma P}{E}\right). \tag{22}$$

The condition for the temperature independence of the WGM radius takes the form:

$$\frac{\sigma P_0}{E} = \alpha'(T - T_0) + \frac{\sigma P}{E}, \tag{23}$$

which, taking (21) in an account, gives:

$$\alpha' + \frac{\sigma}{E}(\alpha' - \alpha_M)\left(\frac{1}{E} + \frac{1}{E'_M}\right)^{-1} = 0. \tag{24}$$

Relation (24) is a condition for the value of $E'_M$, which depends on the material and dimensions of the clamp (see (18)):

$$E'_M = \frac{E}{\sigma(\alpha_M/\alpha' - 1) - 1}. \tag{25}$$

Condition (25) can be satisfied only if the clamp CTE of is significantly higher than that of cylinder. In addition, due to (18), the required value $E'_M$ is less than the elastic modulus $E_M$:

$$E'_M = E_M\left(1 + \frac{Sd^3}{128L_0 ab^3}\right)^{-1} < E_M. \tag{26}$$

For the traditional WGM resonator material - $CaF_2$, which is attractive due to low optical losses, condition (25) is not satisfied, because its CTE $\alpha = 18 \cdot 10^{-6}$ 1/°C does not differ much from that of metals. For quartz cavities, $\alpha = 0.55 \cdot 10^{-6}$ 1/°C, condition (25) can be fully satisfied; paired with an aluminum clamp, we have $E'_M \approx E/10 \sim 10$ GPa. Such a low value of the modulus of elasticity $E'_M$ can be achieved only due to the parameters of the crossbar of the clamp:

$$\frac{Sd^3}{L_0 ab^3} \geq 10^3. \tag{27}$$

This equality can be achieved with a submillimeter thickness of the crossbar $b$. If condition (25) is not satisfied, then the effective CTE can only be reduced to the level:

$$\alpha_{\text{eff}} = \alpha' - \sigma \frac{\alpha_M - \alpha'}{1 + E/E'_M}. \tag{28}$$

**Conclusion**

The presented estimates and numerical calculation show that in the combination of an optical cavity with a stressed mechanical element (ring or clamp), zero CTE can be achieved. The use of stressed elements is most effective for cavities with small CTE values (quartz, silicon). The temperature of zero CTE in the combined device depends on the initial tension of the ring or clamp. Simulations have shown that the fused silica Fabry-Perot cavity, compressed with an aluminum ring, has a point of zero thermal expansion and exhibits approximately half the temperature sensitivity as compared to the temperature-compensated ULE glass cavity (Corning Code 7972). The proposed method of temperature compensation gives a new way to design reference optical cavities with zero thermal expansion without need in materials with zero CTE.

**Funding.** N. O. Zhadnov thanks Russian Science Foundation for funding (grant no. 19-72-10166).